\documentclass[12pt]{elsart}
\usepackage{graphicx,epsfig}
\begin{document}
\begin{frontmatter}
\title{Diquark condensation effects on hot quark star configurations}
\author[rostock,dubna]{D. Blaschke,}
\author[lulea]{S. Fredriksson,}
\author[rostock,yerevan]{H. Grigorian\thanksref{DFG},}
\author[lulea,ankara]{A.M. \"Ozta\c{s}}
\address[rostock]{Fachbereich Physik, Universit\"at Rostock, D-18051 Rostock,
Germany}
\address[dubna]{Bogoliubov  Laboratory of Theoretical Physics,\\
        Joint Institute for Nuclear Research, 141980, Dubna, Russia}
\address[lulea]{Department of Physics, Lule{\aa} University of Technology,
SE-97187 Lule\aa , Sweden}
\address[yerevan]{Department of Physics, Yerevan State University,
375025 Yerevan, Armenia}
\address[ankara]{Department of Physics, Hacettepe University,
TR-06532 Ankara, Turkey}
\thanks[DFG]{Supported by DFG grant No 436 ARM 17/5/01}
\begin{abstract}
The equation of state for quark matter is derived for a nonlocal, chiral quark
model within the mean field approximation.
We investigate the effects of a variation of the formfactors of the
interaction on the phase diagram of quark matter.
Special emphasis is on the occurrence of a diquark condensate which signals a
phase transition to color superconductivity and its effects on the equation of
state  under the condition of $\beta$- equilibrium and charge neutrality.
We calculate the quark star configurations by solving the Tolman- Oppenheimer-
Volkoff equations and obtain for
the transition from a hot, normal quark matter core of a protoneutron star
to a cool diquark condensed one a release of binding energy of the order of
$\Delta M c^2 \sim 10^{53}$ erg.
We find that this energy could not serve as an engine for explosive
phenomena since the phase transition is not first order.
Contrary to naive expectations the mass defect increases when for a given
temperature  we neglect the possibility of diquark condensation.

\noindent PACS number(s):  04.40.Dg, 12.38.Mh, 26.50.+x, 97.60.Bw

\end{abstract}
\end{frontmatter}
\newpage
\section{Introduction}

Color superconductivity (CS) in quark matter \cite{Rajagopal:2000wf} is one
interesting aspect of
the physics of compact star interiors \cite{Blaschke:uj}.
Since calculations of the energy gap of quark pairing predict
a value  $\Delta \sim 100$ MeV and corresponding critical
temperatures for the phase transition to the superconducting state are
expected to follow the BCS relation $T_c = 0.57 ~\Delta$, the question arises
whether diquark condensation can lead to remarkable
effects on the structure and evolution of compact objects.
If positively answered, CS of quark matter could
provide signatures for the detection of a deconfined phase in the
interior of compact objects (pulsars, low-mass X-ray binaries)
along the lines of previously suggested strategies
\cite{Glendenning:1997fy,Glendenning:2001dr,Poghosyan:2000mr,Prakash:2002xx,Grigorian:2002ih}.

Diquarks gained popularity in astrophysics about a decade ago,
when they were suggested to influence the supernova collapse and
``bounce-off'' \cite{fredriksson89,kastor91,Horvath:ms,Horvath:ij,sahu93},
and
to enhance the neutrino cooling of quark-stars. The latter effect
is now subject to much research within improved scenarios Refs.
\cite{Blaschke:1999qx,Page:2000wt,Blaschke:2000dy}.
Possible consequences for the magnetic field and glitch-type phenomena
are discussed in \cite{Blaschke:1999fy,Sedrakian:2000kw,Iida:2002ev}.

Hong, Hsu and Sannino have conjectured \cite{Hong:2001gt}
that the release of binding energy due to Cooper pairing of quarks in
the course of protoneutron star evolution could provide an explanation
for the unknown source of energy in supernovae, hypernovae or gamma-ray
bursts, see also \cite{Ouyed:2001bm}.
Their estimate of energy release was based on the simple formula
\begin{equation}
\Delta E \sim \left(\frac{\Delta}{\mu}\right)^2 M~c^2 \sim 10^{53}~{\rm erg}~,
\end{equation}
which did not take into account the change in the gravitational binding
energy due to the change in the structure of the stars quark core.
In the present work we will reinvestigate the question of a possible binding
energy release due to a CS transition by taking into
account changes in the equation of state (EoS) and the configuration of the
quark star selfconsistently.

As a first step in this direction we will discuss here
the two flavor color superconducting (2SC) quark matter
phase which occurs at lower baryon densities than the
color-flavor-locking (CFL) one, see \cite{Steiner:2002gx,Neumann:2002jm}.
Studies of three-flavor quark models have revealed a very rich phase structure
(see \cite{Neumann:2002jm} and references therein).
However, for applications to compact stars the omission of the strange quark
flavor within the class of nonlocal chiral quark models considered here
may be justified by the fact that central chemical potentials in stable
star configurations do barely reach the threshold value at which the mass gap
for strange quarks breaks down and they appear in the system
\cite{Gocke:2001ri}.
Therefore we will not discuss here first applications to calculate compact star
configurations with color superconducting quark matter phases that have
employed non-dynamical quark models \cite{Lugones:2002zd,Alford:2002rj}.
Their results depend on the somewhat arbitrary choice of the strange quark mass
$m_s$, the diquark gap $\Delta$, and the bag constant.
Dynamical calculations like the one we are going to present here fix their
model parameters with the constraint to describe observable hadron properties.
They include a selfconsistent solution of the chiral and diquark gap equations
which shows that the bulk of a compact star interior does not contain strange
quark matter \cite{Gocke:2001ri,Baldo:2002ju}.

We will investigate the influence of the formfactor of the interaction
on the phase diagram and the EoS of dense quark matter
under the conditions of charge neutrality and
isospin asymmetry due to $\beta$-equilibrium relevant for compact stars.

Finally we consider  the question whether the effect of
diquark condensation which occurs in the earlier stages of the compact star
evolution ($t \simeq 100$ s) \cite{Blaschke:2000dy,carter00,jaikumar01} at 
temperatures $T \sim T_c \sim 20-50$ MeV
can be considered as an engine for exposive astrophysical phenomena like
supernova explosions due to the release of a binding energy of
about $10^{52} \div 10^{53}$ erg, as has been suggested before
\cite{Hong:2001gt,Ouyed:2001bm}.

\section{Thermodynamic potential for asymmetric 2SC quark matter}

We consider a nonlocal chiral quark model described by the effective action
\begin{eqnarray}
S[\bar\psi,\psi]&=&\sum_p\bar\psi(\;\!\!\not\!p-\hat{m})\psi+
S_{q\bar q}[\bar\psi,\psi]+S_{qq}[\bar\psi,\psi],\\
S_{q\bar q}[\bar\psi,\psi]&=&{G_1}\sum_{p,p'}
\left[(\bar\psi g(p) \psi)(\bar\psi g(p') \psi)+
(\bar\psi i\gamma_5\vec{\tau} g(p) \psi)(\bar\psi i\gamma_5\vec{\tau} g(p')
\psi) \right],\\
S_{qq}[\bar\psi,\psi]&=&{G_2}\sum_{p,p'} (\bar\psi
i\gamma_5\tau_2\lambda_2C g(p) \bar\psi^T)(\psi^{T} C
i\gamma_5\tau_2\lambda_2 g(p') \psi),
\end{eqnarray}
which generalizes the approach of Ref. \cite{schmidt94} by including the scalar
diquark pairing interaction channel with a coupling strength $G_2$.
Here, the matrix $C=i\gamma_0\gamma_2$ is the charge conjugation
matrix operator for fermions, the matrices $\tau_2$ and $\lambda_2$ are
Pauli and Gell-Mann matrices acting in, respectively, the SU(2) flavor and
SU(3) color spaces;
$\hat{m}$ is the diagonal current quark mass matrix
$\hat{m}=\mathrm{diag}(m_u,m_d)$.
The function $g(p)$ is the formfactor of the separable four-fermion
interaction.
In the present work it is assumed to be instantaneous, so that it depends only
on the modulus of the three-momentum.
After bosonization and integrating out the quark fields, the mean field
approximation introduces the following order parameters: the diquark gap
$\Delta$, which can be seen as the gain in energy due to diquark condensation,
and the mass gaps $\phi_u$, $\phi_d$, which indicate chiral symmetry
breaking. The grand canonical thermodynamic potential per volume
can be obtained as
\begin{eqnarray}\label{thpot}
\Omega_q(\phi_u,\phi_d,\Delta;\mu_u,\mu_d,T)&=&
-T\sum_n\!\int\! \frac{d^3p}{(2\pi)^3}\;
\frac12 {\rm Tr}\ln\left(\frac{1}{T}\tilde S^{-1}(i\omega_n,\vec{p})\right)
\nonumber\\
&+&\frac{\phi_u^2+\phi_d^2}{8~G_1}+\frac{|\Delta|^2}{4~G_2}~,
\end{eqnarray}
where $\omega_n=(2n+1)\pi T$ are the Matsubara frequencies for fermions.
The chemical potential matrix $\hat{\mu}=\mathrm{diag}(\mu_u,\mu_d)$ as
well as the dynamical quark mass matrix
$\hat{M}=\mathrm{diag}(m_u+\phi_u g(p),m_d+\phi_d g(p))$
are included into the definition of the inverse quark propagator
\begin{equation}
\tilde S^{-1}(p_0,\vec{p})=
\left(
\begin{array}{cc}
\not\!p - \hat M -\hat{\mu}\gamma_0~~~&
\Delta \gamma_5\tau_2\lambda_2 g(p)\\
-\Delta^\ast \gamma_5\tau_2\lambda_2 g(p)&
\not\!p - \hat M+\hat{\mu}\gamma_0
\end{array}
\right)~.
\end{equation}
One can combine the chemical potentials $\mu_{u}$, $\mu_{d}$ of $u$ and
$d$ quarks by introducing
$\mu_{q} = (\mu_{u} + \mu_{d})/2$
and $\mu_{I} = (\mu_{u} - \mu_{d})/2$ as the Lagrange multipliers related to,
respectively, the quark number density $n_q$ and the isospin asymmetry $n_I$.
One can easily check the relations
$n_{q} = n_{u} + n_{d} = -{\partial \Omega}/{\partial
\mu_{q}}$, $n_{I} = n_{u} - n_{d} = -{\partial \Omega}/{\partial
\mu_{I}}$.
Following Ref. \cite{Frank:2003ve} we use in the following that the chiral 
gaps of both flavors are degenerate
$ \phi_u = \phi_d = \phi$ (and therefore also the masses $m_u = m_d = m$).
The trace operations in the bispinor, flavor and color spaces as well as the
Matsubara summations can be performed with the following result for the
thermodynamic potential of asymmetric quark matter
\cite{Blaschke:2001ew,kiriyama01,Huang:2002zd}
\begin{eqnarray}\label{ThP}
& &  \Omega_q(\phi,\Delta;\mu_q,\mu_I,T) =
  \frac{\phi^2}{4~G_1}+\frac{\Delta^2}{4~G_2}
    - 2 \int^\infty_0\frac{q^{2}~dq}{2 \pi^2}\bigg\{
  2~E_\phi  \nonumber\\
& &  + T~\ln\left[1 +\exp\left(-\frac{E_\phi - \mu_q -\mu_I}{T}\right)\right]
  + T~\ln\left[1 +\exp\left(-\frac{E_\phi - \mu_q +\mu_I}{T}\right)\right]
 \nonumber \\
& & + T~\ln\left[1+\exp\left(-\frac{E_\phi + \mu_q -\mu_I}{T}\right)\right]
  + T~\ln\left[1+\exp\left(-\frac{E_\phi + \mu_q +\mu_I}{T}\right)\right]
\bigg\} \nonumber\\
& &  - 4 \int^\infty_0\frac{q^{2}~dq}{2 \pi^2}\bigg\{
  E_{+} + E_{-} \nonumber\\
& & + T~\ln\left[1+\exp\left(-\frac{E_-  -\mu_I}{T}\right)\right]
  + T~\ln\left[1+\exp\left(-\frac{E_- +\mu_I}{T}\right)\right]
 \nonumber \\
& &+ T~\ln\left[1+\exp\left(-\frac{E_+ -\mu_I}{T}\right)\right]
  + T~\ln\left[1+\exp\left(-\frac{E_+ +\mu_I}{T}\right)\right]\bigg\}\\
\nonumber
& &- \Omega_q^{\rm vac}~.
\end{eqnarray}
The number of color degrees of freedom in our model is fixed to be $N_c = 3$.
Only the coupling constant $G_1$
is considered to be a free parameter of the model,
while $G_2$ could be chosen according to Ref. \cite{berges99} as
$G_2 = G_1~N_c/(2 N_c - 2)$. We discuss later the special choices
$G_2=0.75~G_1$ and $G_2=G_1$.

The dispersion relation $E_\phi = \sqrt{q^2 + (m + \phi ~g(q))^2}$ holds for
unpaired quarks also in color-symmetry broken 2SC phase.
The quarks with paired colors have a modified dispersion relation
\[
E_\pm = (E_\phi \pm
\mu_q)\sqrt{1 + \frac{\Delta^2 ~g^2(q)}{(E_\phi \pm
 \mu_q)^2}}~.
\]

For nonvanishing $\Delta$ in Eq. (\ref{ThP}) the color symmetry is broken.
Two of the three quark color degrees of freedom are
coupled to bosonic Cooper pairs in the color antitriplet state which can form
a Bose condensate.
The dispersion relation for quarks of these colors contains the diquark gap
$\Delta$ and becomes
\begin{equation}
E_{f\pm}^2 = (E_{\phi_f} \pm \mu_f)^2 +\Delta^2 ~g^2(q)~.
\end{equation}
For quarks of the remaining third color there is no pairing gap in
their dispersion relation which reads
\begin{equation}
E_{\phi_f} = \sqrt{q^2 + (m_f + \phi_f ~g(q))^2}~.
\end{equation}
In thermal equilibrium characterized by a fixed set of thermodynamic variables
$\{\mu_{u}, \mu_{d}, T\}$, the thermodynamic potential $\Omega_q$ shall attain
a global minimum with respect to all its order parameters.
In the corresponding state the gap equations for the order parameters
are fulfilled. However, the opposite is not always true. Satisfying
the gap equations is a necessary, but not a sufficient condition for the
thermodynamic equilibrium state.

In Eq. (\ref{ThP}), we subtract the term $\Omega_q^{\rm vac}$ in
order to define the thermodynamic potential such that the vacuum pressure
vanishes,
\begin{equation}
P_q(0,0,0)=0~.
\end{equation}
On the mean-field level the present approach is analogous to a flavor-
asymmetric generalization of the instanton-motivated model of Ref.
\cite{berges99} for two-flavor color superconductivity (2SC).

From BCS theory it is known, that in order to form Cooper pairs at $T=0$ in a
dense Fermi system, the difference in the chemical potentials of the Fermions
to be paired should not exceed the size of the gap.
As previous calculations within this type of models have shown \cite{berges99},
there is a critical chemical potential for the occurrence of quark matter
$\mu_f\ge 300$ MeV and values of the gap in the region $\Delta \le 150$ MeV
have been found.
Therefore it is natural to consider the problem of the color superconducting
(2SC) phase with the assumption, that quark matter is symmetric or very
close to being symmetric ($\mu_u \simeq \mu_d$).

The number densities derived
from thermodynamic potential can be expressed in an integral form as
\begin{eqnarray}
   n_q(\mu_q,\mu_I,T) &=&  \nonumber
2 \int^\infty_0\frac{q^{2}~dq}{2 \pi^2}\left\{f(E_\phi-\mu_q-\mu_I,T) +
   f(E_\phi-\mu_q+\mu_I,T)\right.\\ \nonumber
& &    \left.-f(E_\phi+\mu_q-\mu_I,T)
- f(E_\phi+\mu_q+\mu_I,T)\right.\\ \nonumber
& &    \left.- 2~{\partial E_+}/{\partial \mu_q}\left
[f(E_+-\mu_I,T)
+ f(E_++\mu_I,T) - 1\right]\right.\\
& &    \left. - 2~{\partial E_-}/{\partial \mu_q}\left[f(E_--\mu_I,T)
+ f(E_-+\mu_I,T) - 1\right]\right\}
\end{eqnarray}
and
\begin{eqnarray}
   n_I(\mu_q,\mu_I,T)& =& \nonumber
2 \int^\infty_0\frac{q^{2}~dq}{2 \pi^2}\left\{f(E_\phi-\mu_q-\mu_I,T)
   - f(E_\phi-\mu_q+\mu_I,T)\right.\\ \nonumber
& &    \left.+ f(E_\phi+\mu_q-\mu_I,T)
- f(E_\phi+\mu_q+\mu_I,T)\right.\\ \nonumber
& &   \left. + 2~\left[f(E_+-\mu_I,T) - f(E_++\mu_I,T)\right.\right.\\
& &   \left. + \left.f(E_--\mu_I,T) - f(E_-+\mu_I,T)\right]\right\}~.
\end{eqnarray}
Here,
$f(E,T) = [1+\exp(E/T)]^{-1}$
is the Fermi distribution function and the derivatives are given by
$\partial E_{\pm}/\partial \mu_q = (\mu_q \pm E_\phi)/E_\pm$.

Using the entropy density $s_q = - \partial \Omega_q/ \partial T$, the
pressure $P_q=-\Omega_q$
and the thermodynamic identity $P_q+\varepsilon_q =s_qT+\mu_q n_q +\mu_In_I$,
the EoS for the energy density as a function of temperature and
chemical potentials becomes
\begin{eqnarray}
\varepsilon(\mu_q,\mu_I,T) &=&  \nonumber
2 \int^\infty_0\frac{q^{2}dq}{2 \pi^2}\left\{ E_\phi
\left[f(E_\phi-\mu_q-\mu_I,T) +
f(E_\phi-\mu_q+\mu_I,T)\right.\right.\\ \nonumber
& &   \left. \left.+ f(E_\phi+\mu_q-\mu_I,T)
 + f(E_\phi+\mu_q+\mu_I,T) -2 \right]\right.\\ \nonumber
& &   \left.+ 2~\left(E_+ - \mu_q~{\partial E_+}/{\partial \mu_q}\right)
\left[f(E_+-\mu_I,T)  + f(E_++\mu_I,T) - 1\right]\right.\\ \nonumber
& &   \left.+ 2~\left(E_- - \mu_q~{\partial E_-}/{\partial \mu_q}\right)
\left[f(E_--\mu_I,T) + f(E_-+\mu_I,T) - 1\right]\right\}\\
& &+ \frac{\phi^2}{4~G_1}+\frac{\Delta^2}{4~G_2} - \Omega_{\rm vac}~.
   \end{eqnarray}

In the next sections we will apply this model for the construction of
the EoS for stellar matter in $\beta$-equilibrium.

\section{EoS for 2SC quark matter in $\beta$-equilibrium }

We consider stellar matter in the quark core of compact stars
consisting of electrons in chemical equilibrium with $u$ and $d$
quarks. Hence $\mu_d = \mu_u + \mu_e$, where $\mu_e$ is the
electron chemical potential.
The thermodynamic potential of such matter is
\begin{equation}
\label{ThPBeta}
  \Omega(\phi,\Delta;\mu_q,\mu_I,\mu_e,T) =
  \Omega_q(\phi,\Delta;\mu_q,\mu_I,T)+\Omega_e(\mu_e,T),
\end{equation}
with
\begin{equation} \Omega_e(\mu_e,T) = -
\frac{1}{12\pi^{2}}\mu_{e}^{4}
-\frac{1}{6}\mu_e^{2}T^{2}-\frac{7}{180}\pi^{2}T^{4}
\end{equation}
for ultrarelativistic electrons.
The number density of electrons is then $n_{e} = - {\partial
\Omega}/{\partial \mu_{e}}$.
The condition  of local charge neutrality
\begin{equation}
\frac{2}{3}n_u - \frac{1}{3}n_d - n_e = 0,
\end{equation}
in the new notation reads
\begin{equation}
n_{q} + 3n_{I} - 6n_{e} = 0.
\end{equation}

In our model we assume the neutrino chemical potential to be zero, which
corresponds to a scenario where the neutrinos (and/or antineutrinos)
leave the star as soon as they are created.
The limits of such an approximation could be based on the
calculations for the case of trapping of neutrinos in the hot stars interior
\cite{Steiner:2002gx}.
For possible consequences of antineutrino trapping on hot quark star
evolution, see  \cite{Aguilera:2002dh}.

In \cite{bedaque02,kiriyama01} it is
shown that at given $\mu_q$ the diquark gap is independent of the
isospin chemical potential for $|\mu_{I}(\mu_q)|<\mu_{Ic}(\mu_q)$,
otherwise  vanishes.
Increase of isospin asymmetry forces the system to undergo
a first order phase transition by tunneling through a
barrier in the thermodynamic potential (\ref{ThPBeta}).
In order to describe this transition we have to choose the 
apropriate set of Lagrange multipliers. Under the constraints of 
$\beta-$equilibrium and charge neutrality,   
the free enthalpy density can be written in the form
\begin{equation}
G=\mu_u~n_u + \mu_d~n_d + \mu_e~n_e = \mu_B~n_B~,
\end{equation}
where $\mu_B=\mu_u+2\mu_d$ is the chemical potential conjugate to the
baryon number density $n_B=(n_u+n_d)/3$ and remains as an independent
thermodynamical variable under the given constraints.
Note that $\mu_B\neq \mu_q/3$ because of the isospin violation in weak 
interactions, which determine the equilibrium conditions for compact stars.
This is in contrast to equilibrium conditions for dense matter in 
heavy-ion collisions.
The corresponding set of equations of state is obtained from the thermodynamic
potential (\ref{ThPBeta}) in the form
\begin{eqnarray}
P(T,\mu_B)&=&-\Omega(T,\mu_B)~,\\
n_B(T,\mu_B)&=&-{\partial \Omega(T,\mu_B)}/{\partial \mu_B}~,\\
s(T,\mu_B)&=&{\partial \Omega(T,\mu_B)}/{\partial T}~,\\
\varepsilon(T,\mu_B)&=&T ~s(T,\mu_B)-P(T,\mu_B)+\mu_B~n_B(T,\mu_B)~,
\end{eqnarray}
where due to the above constraints the dependence on the set of 
thermodynamic potentials 
$(\mu_q,\mu_I,\mu_e)$ is reduced to the dependence on $\mu_B$ only.

In concluding this section we would like to comment on the order of
the CS phase transition within above approach.
The phase transition to the 2SC phase coming from
the hot normal state of matter by cooling the system
occurs as a second order phase transition without a jump of
concentrations, whereas the phase transition from
an isospin asymmetric state to a superconducting one by
decreasing the asymmetry is of first order and has a jump in the densities
(in particular in the electron fraction).

\section{Results of model calculations}

\subsection{EoS of quark matter in 2SC phase for finite temperature}

\begin{figure}[bth]
  \begin{center}
    \includegraphics[width=0.8\textwidth,angle = -90]{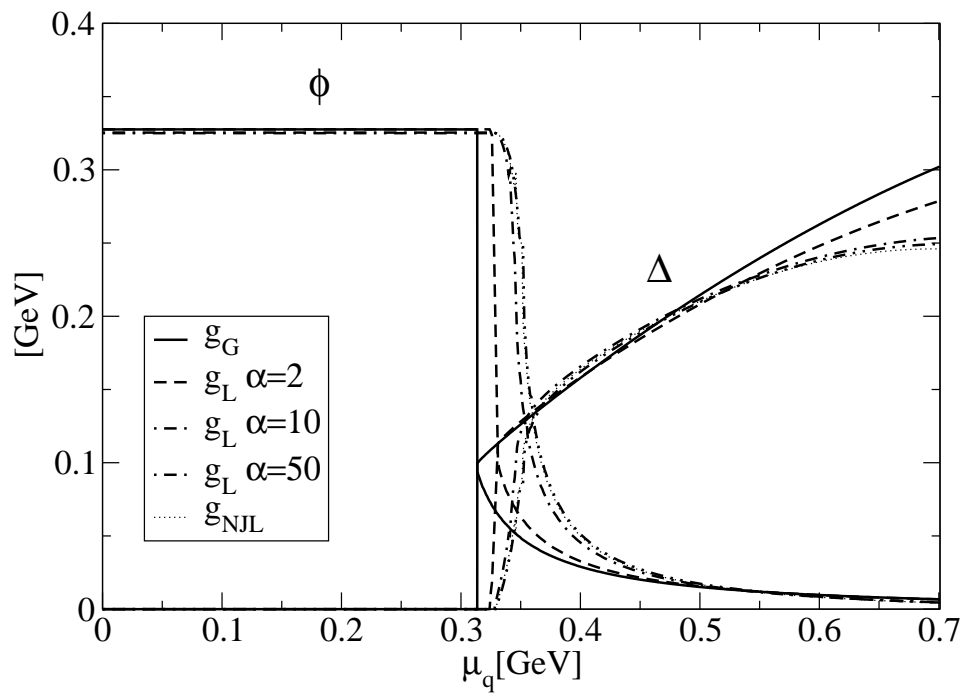}
    \caption{Solutions for the chiral gap $\phi$ and the diquark gap $\Delta$
 for different formfactors at $T=0$ in $\beta-$equilibrium quark matter, 
$G_2=G_1$.}
    \label{fig1}
  \end{center}
\end{figure}

\begin{figure}[bth]
  \begin{center}
    \includegraphics[width=0.8\textwidth,angle =-90]{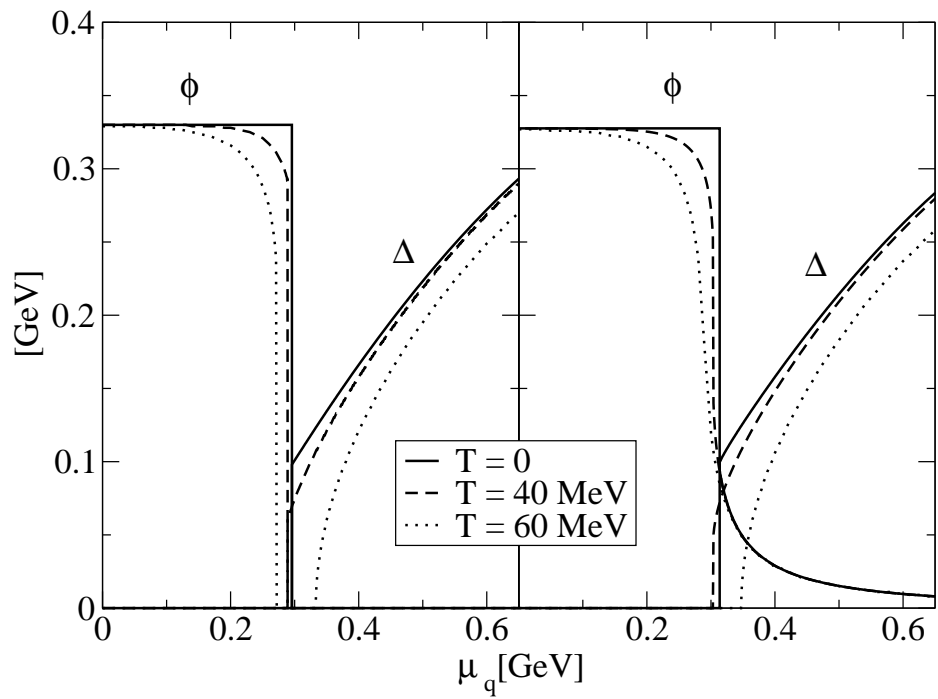}
    \vspace*{0.3cm}
    \caption{Solutions of the chiral gap $\phi$ and the diquark gap $\Delta$
 for the Gaussian model formfactor
in the chiral limit (left panel) and for finite current quark mass
$m_0 = 2.41$ MeV (right panel)
at different temperatures $T=0,40,60$ MeV for $\beta-$equilibrium quark matter,
$G_2=G_1$.}
    \label{fig2}
  \end{center}
\end{figure}

The gaps have been calculated by minimizing the thermodynamic
potential $\Omega$, Eq. (\ref{ThPBeta}), in the space of order parameters
($\phi,\Delta$)  and the
results are shown in Figs. \ref{fig1} and \ref{fig2}.

In order to compare the effects of the quark interaction formfactors on
the EoS, we study  the Gaussian (G), Lorentzian (L) and cutoff (NJL) type
formfactors defined as
\begin{eqnarray}
\label{LF}
g_{\rm L}(q) &=& [1 + (q/\Lambda_{\rm L})^{2\alpha}]^{-1},~~~~ \alpha > 1~,
\\
\label{GF}
g_{\rm G}(q) &=& \exp(-q^2/\Lambda_{\rm G}^2)~,
\\
\label{NF}
g_{\rm NJL}(q) &=& \theta(1 - q/\Lambda_{\rm NJL})~.
\end{eqnarray}
The Lorentzian (L) interpolates between a soft
(Gaussian-type, $\alpha \sim 2$) and a
hard cutoff (NJL, $\alpha > 30$) depending on the value of the parameter
$\alpha$.

We will employ parametrisations of the
nonlocal quark model which reproduce pion properties: mass $m_\pi=140$ MeV,
decay constant $f_\pi=93$ MeV and which have the same quark mass gap
$\phi(T = 0,\mu= 0) = 330$ MeV in the vacuum. The
results for the parameterization are taken from \cite{schmidt94},
see Tab. \ref{par}. $G_2$ remains a free parameter of the approach.
For $G_2=G_1$, the 2SC phase with nonvanishing diquark condensate is obtained
whereas for $G_2=0.75~G_1$ no 2SC phase is possible under neutron star 
conditions.

Figure \ref{fig1} displays the $T=0$ solutions of the chiral gap
$\phi$ and diquark gap $\Delta$ for different formfactors in 
$\beta-$equilibrium for $G_2=G_1$.
For the densities relevant for stable star configurations,
$\mu_{q}\le 450$ MeV, the critcal chemical potential $\mu_q^c$
for the chiral transition and for the onset of diquark condensatuion
does depend on the type of the formfactor. The value of
the diquark gap reaches $\Delta \sim 200$ MeV in the range of densities
relevant for compact stars. It is rather insensitive to changes of the 
formfactor of the quark interaction.

\begin{figure}[bth]
  \begin{center}
    \includegraphics[width=0.8\textwidth,angle =-90]{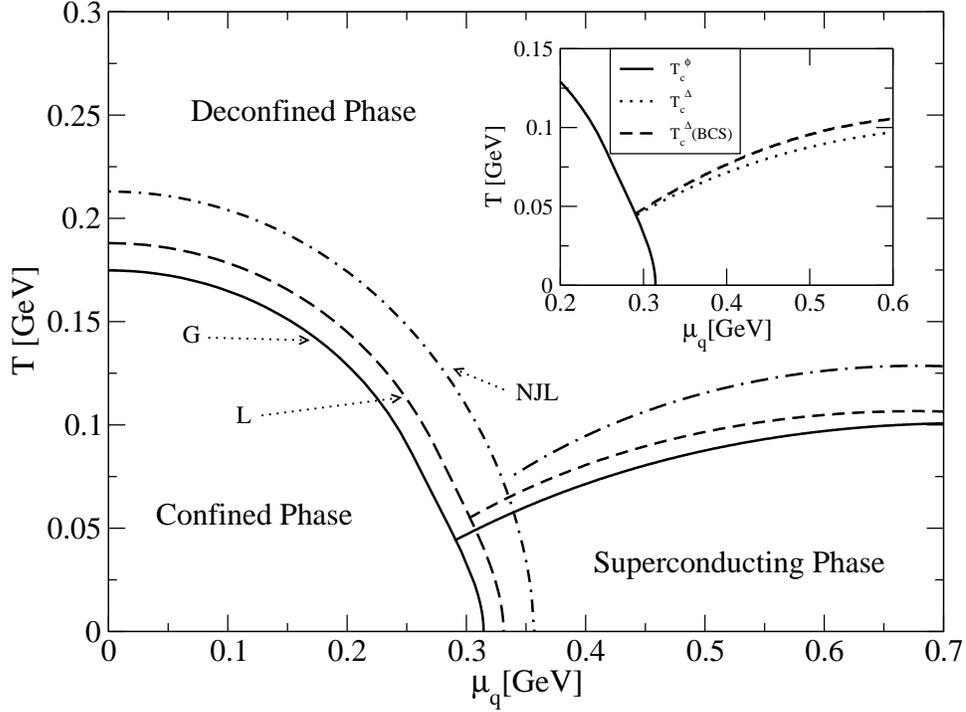}
    \caption{Phase diagrams for different
formfactor models: Gaussian (solid lines), Lorentzian $\alpha=2$ (dashed
lines)  and NJL (dash-dotted) under conditions of
$\beta$-equilibrium and charge neutrality. 
The color-superconducting phase which exists for $G_2=G_1$ is absent 
for  $G_2=0.75~G_1$.
In the inset we show for the Gaussian
model the comparison of the numerical result with the modified
BCS formula  $T_c^{\Delta} = 0.57~ \Delta(T=0,\mu_q)~g(\mu_q)$
for the critical temperature of the superconducting phase transition.
}
    \label{fig3}
  \end{center}
\end{figure}

Figure \ref{fig2} shows the solutions for the chiral gap $\phi$ and the
diquark gap $\Delta$
for the Gaussian model formfactor at different temperatures $T=0, 40, 60$ MeV.
We compare results in the chiral limit ($m_0=0$) with those for finite
current quark mass $m_0 = 2.41$ MeV and observe that the diquark gap is not
sensitive to the presence of the current quark mass,
which holds for all formfactors in Eqs. (\ref{LF})- (\ref{NF}).
However, the choice of the formfactor influences the critical
values of the phase transition as displayed in the quark matter phase diagram
($\mu_{q} - T$ plane) of Fig. \ref{fig3}, see also Fig. \ref{fig1}.
A softer formfactor in momentum space gives lower critical values for
$T_c$ and $\mu_c$ at the borders of chiral symmetry restoration and diquark
condensation.
The numerical results for the critical chemical potentials at $T = 0$ are
summarised in Tab. \ref{par} for the different formfactors.

The inset of Fig. \ref{fig3} shows that the generalisation of the BCS relation
$T_c \simeq 0.57~\Delta(T=0,\mu_q)~g(\mu_q)$,
between the critical temperature $T_c$ of the superconducting phase transition
and the pairing gap $\Delta$ at $T=0$ is satisfactorily fulfilled in the
domain of the phase diagram relevant for compact stars.

\begin{figure}[bth]
  \begin{center}
    \includegraphics[width=0.8\textwidth,angle = -90]{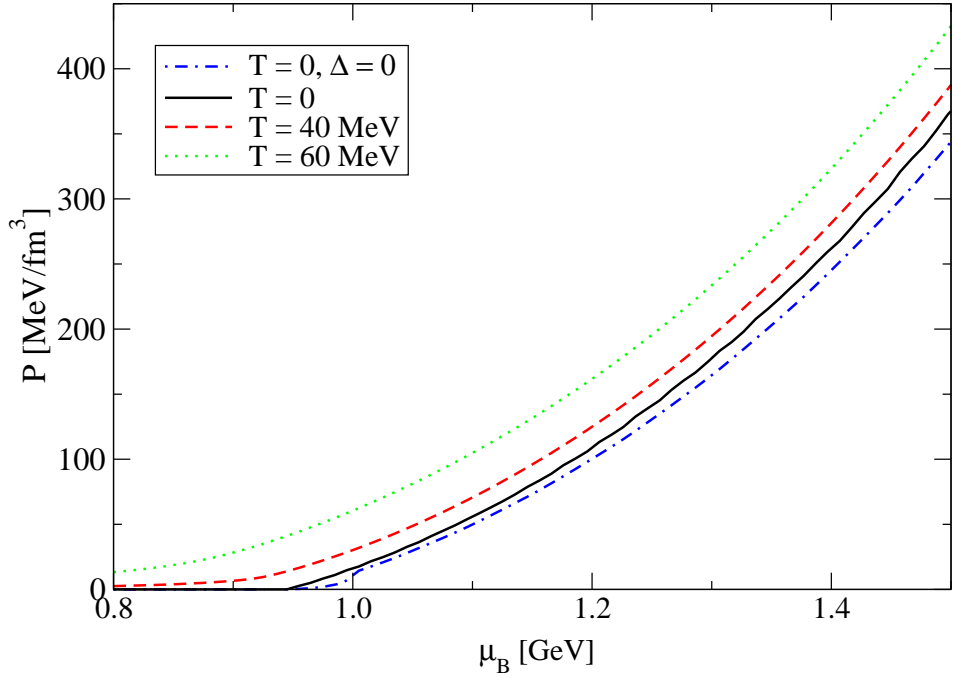}
    \caption{Dependence of the pressure on the baryochemical $\mu_B$ of
quark matter for the Gaussian formfactor model for different temperatures
$T=0,40,60$ MeV. To demonstrate the effect of the stiffening of the EoS
due to diquark condensation, we show the $T=0$ result for both $\Delta \neq 0$
(solid line, $G_2 = G_1$ ) and $\Delta = 0$
(dash-dotted line, $G_2=0.75~G_1$).}
    \label{fig4}
  \end{center}
\end{figure}

Figure \ref{fig4} shows the resulting equation of state (EoS) for
quark matter, as pressure $P$ versus baryochemical potential $\mu_B$
for the isotherms $T = 0, 40, 60 $ MeV and in $\beta$-equilibrium.
The Gaussian formfactor has been used.
The Figure displays the defect of nonconfining quark models: at
nonzero temperature quasifree quarks contribute to the
pressure at low chemical potentials $\mu_B$ 
(corresponding to energy densities $\varepsilon \le 300$ MeV/fm$^3$).
The dot-dashed curve in  Fig. \ref{fig4} shows that the presence of
a diquark condensate stabilizes the quark matter system and entails
a hardening of the EoS.

\begin{figure}[bth]
  \begin{center}
    \includegraphics[width=0.8\textwidth,angle = -90]{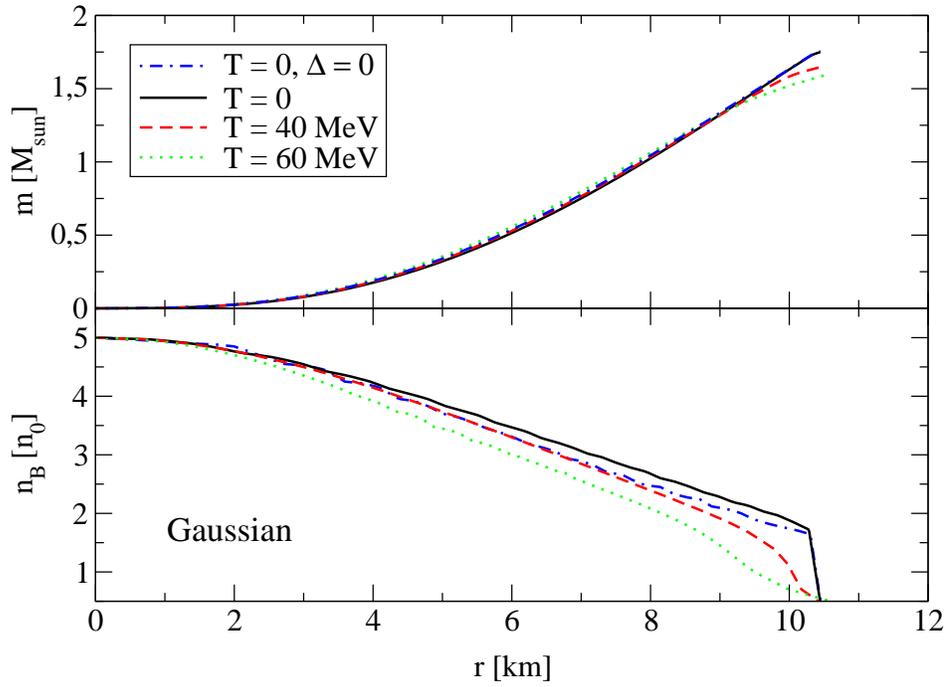}
    \caption{Matter distribution inside the quark star
(lower panel) and accumulated mass (upper panel) for
    the quark stars with the same central baryon density
$n_B(0) = 5~ n_0$ and cut on $n_B(R) = 0.5~n_0$,
    for different temperatures.
The case $\Delta=0$ is calculated for  $G_2=0.75~G_1$, the others for
$G_2 = G_1$.}
    \label{fig5}
  \end{center}
\end{figure}

In Figure \ref{fig5} we show the matter distribution inside the quark 
star and the accumulated mass for quark stars with the same central 
baryon density $n_B(0) = 5~ n_0$ for different temperatures.
We have cut the density distribution at finite temperatures 
for $n_B < 0.5 ~n_0$ due to the following reason.
At nonzero temperatures the mass gap decreases as a function of the
chemical potential already in the phase with broken chiral symmetry.
Hence the model here gives unphysical low-density excitations of quasi-free
quarks.
A systematic improvement of this situation should be obtained by including
the phase transition construction to  hadronic matter.
However, in the present work we circumvent the confinement problem by
considering the quark matter phase only for densities  $n_B > 0.5 ~n_0$,
where $n_0$ is nuclear saturation density.

\subsection{Configurations of hot quark stars}

Here  we compare configurations with and without CS,
in order to investigate the effect of diquark condensation on the total
energy of a quark star and to decide whether the corresponding phase
transition could serve as an engine for explosive phenomena like supernov{\ae}
and gamma ray bursts.
The CS transition can occur during the cooling evolution of a hot protoneutron
star \cite{Blaschke:2000dy,carter00}.
We suppose that the approximation of an isothermal temperature distribution
\cite{Kettner:1994zs} in the star interior is sufficient for the estimate of
the mass defect \footnote{For a discussion of isentropic hot quark star
configurations, see \cite{Blaschke:1998hy}.}.
In the present paper we consider selfbound configurations of pure quark matter
as models for quark cores in hybrid stars which should be the general case to
be discussed in a separate paper, see e.g. Ref. \cite{Grigorian:2003vi} for
$T=0$ configurations.
The extrapolation from pure quark star configurations to hybrid ones,
however, is strongly model dependent and cannot be done without a calculation.
Quark matter effects are expected to be still valid in
hybrid stars when rescaled with a monotonously rising function of the ratio
of the quark core volume to the quark star volume.

The spherically symmetric, static star configurations are defined by the
well known Tolman-Oppenheimer-Volkoff equations
\cite{oppenheimer39} for the mechanical equilibrium of self-gravitating
matter (see also \cite{Blaschke:1998hy,glendenning00})
\begin{equation}
\frac{dP(r)}{dr}= -\frac{[\varepsilon(r)+P(r)][m(r)+4\pi
r^{3}P(r)]}{r[r-2m(r)]}~.
\end{equation}

Here $\varepsilon(r)$ is the energy density and $P(r)$ the pressure at the
distance $r$ from the center of the star. The mass enclosed in a sphere with
radius $r$ is defined by
\begin{equation}
m(r)=4\pi \int_{0}^{r}\varepsilon(r')r'^{2}dr'~.
\end{equation}

\begin{figure}[bth]
  \begin{center}
    \includegraphics[width=0.8\textwidth,angle = -90]{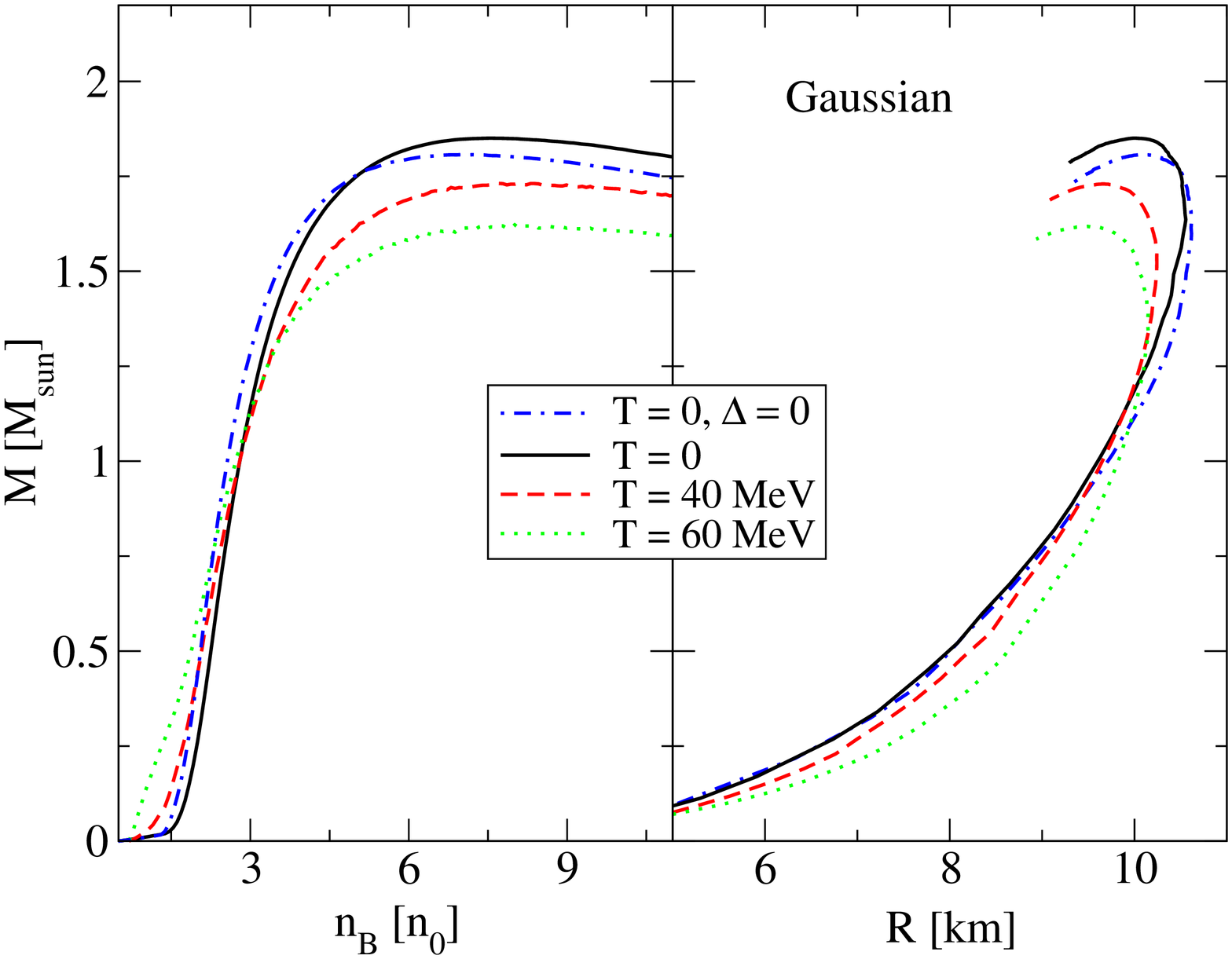}
   \vspace*{-1.5cm}

    \includegraphics[width=0.8\textwidth,angle = -90]{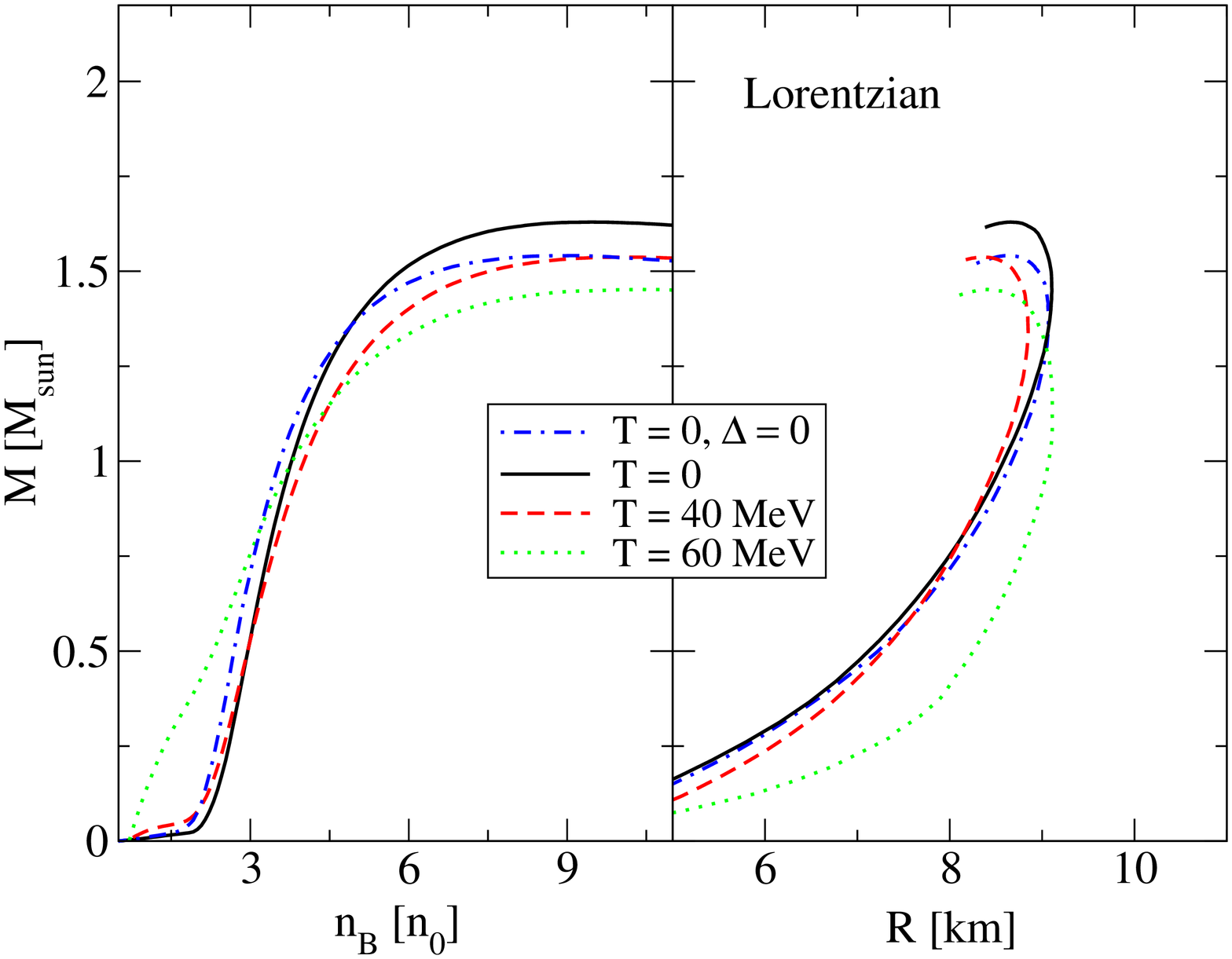}
   \vspace*{-1cm}

    \caption{Stable configurations of quark stars
    for different temperatures $T = 0,40,60$ MeV.
     Mass as a function of central baryon density (left panels) and of the
radius (right panels) for the Gaussian model (upper graph) and for the
Lorentzian model (lower graph).
The case $\Delta=0$ is calculated for  $G_2=0.75~G_1$, the others for
$G_2 = G_1$.}
    \label{fig4c}
  \end{center}
\end{figure}

These equations are solved with the boundary condition of a given central
baryon number density $n_B(0)$ defining a configuration.
In Fig. \ref{fig4c} we show the results for the dependence of the masses on
the central energy density as well as on the radius of the configurations
for different temperatures and two formfactor models: a Gaussian and
a Lorentzain one.
Stable configurations correspond to the rising branch on the mass-central
density plane.
In Tab. \ref{mass} we give for fixed temperatures $T=0,~40,~60$ MeV the
values of maximal possible mass, corresponding radius, central baryon
number density and energy density for a Gaussian  and a
Lorentzian formfactor, respectively.
For comparison, we have included $T=0$ configurations for a small coupling
constant $G_2 =0.75~ G_1$, for which the diquark condensate does not occur.
The comparison shows that the maximum mass decreases with increasing
temperature due to a softening of the EoS.
Also the disappearance of the diquark condensate (in our case due to
lowering of the coupling $G_2$) softens the EoS and lowers the maximum mass.
The Gaussian formfactor has the lowest critical baryochemical potential
when compared with the Lorentzian and the NJL model.
Therefore the Gaussian quark star configuration can extend to larger radii
where the density is lower and can thus carry more mass than the
Lorentzian model with the same central baryon density, see Fig.\ref{fig4c}.

Since in our calculations we do not consider the existence of a possible
hadronic shell of the star where the transition
to the phase with broken chiral symmetry and confinement occurs,
we introduce a phenomenological  cutoff in
the baryon number density at the value $n_B = 0.5 ~n_0$,  see Fig \ref{fig5}.
With this condition we define the radius of the quark core $R$.
The total mass $M$ of the object is defined by the sphere with that radius
$M = m(R)$. It is easy to see from that Figure that temperature plays no role
for the mechanical equilibrium of the quark core close to the center of the
configuration due to its high density.
It has an influence, however, on the peripheral region and changes the total
mass for a given central density, see Figs. \ref{fig4c}.
We have checked numerically, that a change of the somewhat arbitrary cutoff
density does not change the
qualitative result for the energy release reported below.

\begin{figure}[thb]
  \begin{center}
    \includegraphics[width=0.85\textwidth,angle = -90]{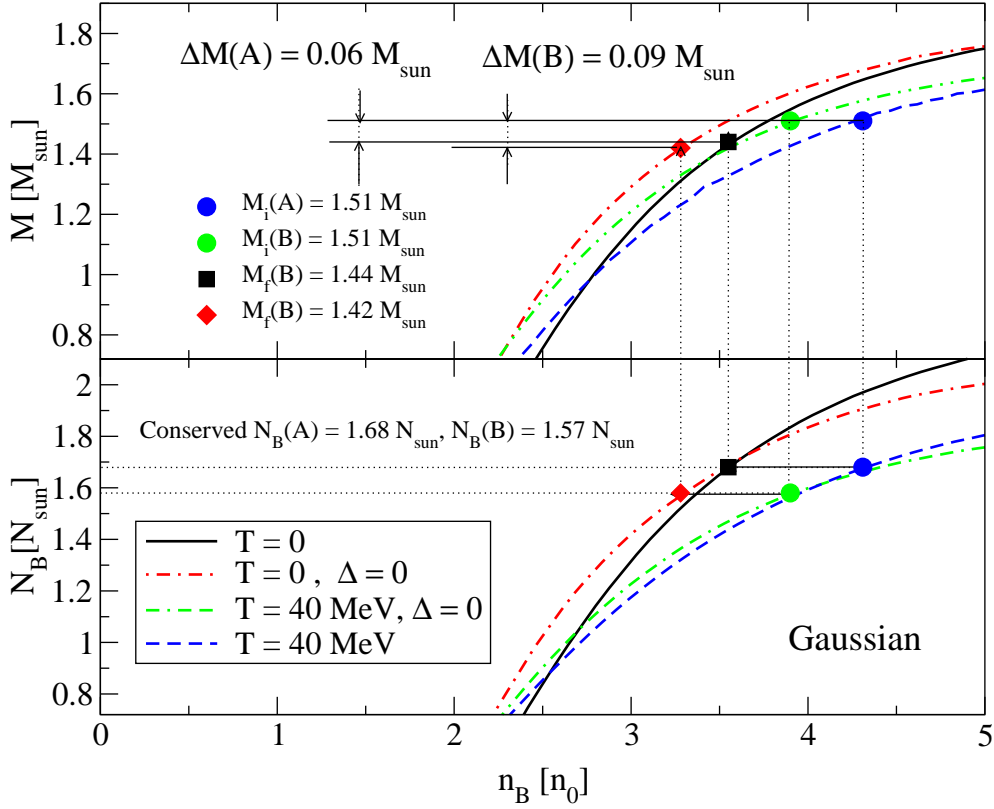}
    \caption{Hot ($T=40$ MeV) versus cold  ($T=0$) quark star configurations
for the Gaussian model;
case A: with diquark condensation (dashed versus full lines)  and
case B: without diquark condensation
(dash-dash-dotted versus dash-dotted lines).
When a quark star with initial mass $M_i$ cools down
from $T=40$ MeV to $T=0$ at fixed baryon
number $N_B$ the mass defect $\Delta M$ occurs.
The case $\Delta=0$ is calculated for  $G_2=0.75~G_1$, the others for
$G_2 = G_1$.}
    \label{fig6}
  \end{center}
\end{figure}

\begin{figure}[thb]
  \begin{center}
    \includegraphics[width=0.85\textwidth,angle = -90]{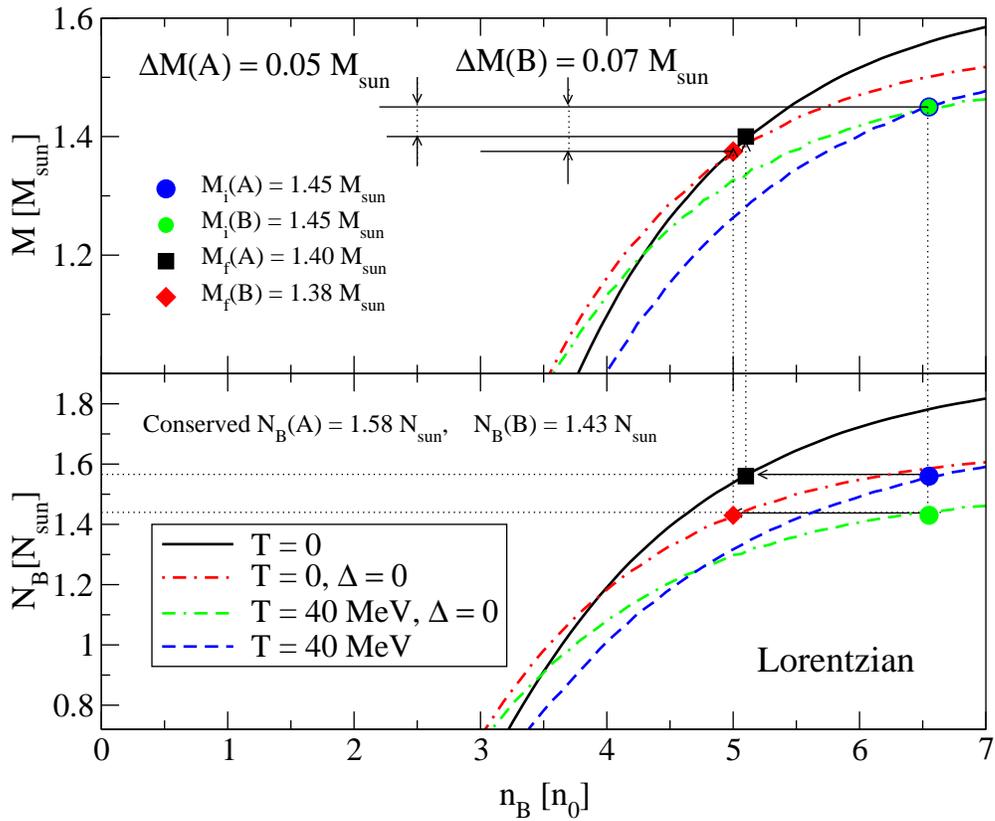}
    \caption{Same as Fig. \ref{fig6} for the Lorentzian model.}
    \label{fig7}
  \end{center}
\end{figure}

For the discussion of evolutionary variations of the internal structure during
the cooling it is  important to take into account the conservation of the
total baryon number of the star.
Therefore, one has to compare configurations with different central densities,
but with the same baryon number $N_B$, see Figs. \ref{fig6}, \ref{fig7}.

Here we will discuss two scenarios for the protoneutron star 
cooling which we denote by  $A$ and $B$, where $A$ stands for cooling of 
a star configuration with SC whereas $B$ is a scenario without SC. 
The initial states for both scenarios are chosen to have the same mass 
$M_i(A)=M_i(B)$ for a given initial temperature of $T=40$ MeV.
The final states at $T=0$, however, have different masses 
$M_f(A)\neq M_f(B)$ while
the total baryon number is conserved in the cooling evolution.
The resulting mass differences 
are $\Delta M(A) = 0.06 ~M_{\odot}$, $\Delta M(B) = 0.09
~M_{\odot}$  and $\Delta M(A) = 0.05 ~M_{\odot}$, $\Delta M(B)
= 0.07 ~M_{\odot}$ for the Gaussian and Lorentzian models, respectively. 

We make the observation that, contrary to naive expectations
of the mass defect 
as a result of the binding energy from Cooper pairing \cite{Hong:2001gt}, 
$\Delta M~c^2\sim (\Delta/\mu)^2 M_{\odot}~c^2 \propto 10^{53}$ erg,
the mass defect in the cooling evolution with diquark condensation (A) is
about $40\%$ smaller than in the scenario without SC (B), regardless
of the choice of formfactor.
The difference amounts to $\simeq 0.02 M_{\odot}$ for
initial configurations with masses of about $1.4-1.5~ M_\odot$.
Therefore, we conclude that general
relativity effects play an important role
in the estimate of the mass defect since the effects of diquark condensation
on the mass of configurations via the change in the equation of state 
overcompensate effects due to the occurence of the diquark gap in the
one-particle energies.

\section{Conclusion}

We have investigated the influence of diquark condensation on
the thermodynamics of quark matter under the conditions of
$\beta$-equilibrium and charge neutrality relevant for the 
discussion of compact stars.
The EoS has been derived for a nonlocal chiral quark model
in the mean field approximation,
and the influence of different formfactors of the nonlocal,
separable interaction (Gaussian, Lorentzian, NJL) has been studied.
The model parameters are chosen such that the same set of hadronic
vacuum observables is described.
We have shown that the critical temperatures and chemical potentials 
for the onset of the chiral and the superconducting phase transition 
are the lower the smoother the momentum dependence of the 
interaction formfactor is. 

The phase transition to color superconducting quark matter from
the lower density regions at small temperatures ($T < 5 \div 10$
MeV) is of first order, while the melting of the diquark
condensate and the corresponding transition to normal quark matter
at high temperatures is of second order. 
The flavor asymmetry in neutral quark matter increases when the electron
fraction drops. 

When the diquark coupling constant is large enough 
($G_2\stackrel{>}{\sim} G_1$) 
then the asymmetry which is present in the system due to the 
$\beta$-equilibrium condition does not destroy the diquark condensation. 
The electron fraction under these conditions does not exceed 
$n_e/n_{\rm total} \simeq 0.01$. 
The masses of the quark core
configurations could be up to $1.8 ~M_{\odot}$ for the Gaussian
and up to $1.6 ~M_{\odot}$ for the Lorentzian formfactor models.
The quark core radii with Lorentzian formfactors are up to $9$ km, 
while for Gaussian models could extend up to $11$ km.

Diquark condensation makes the EoS harder, which leads to an
increase in the maximum mass of the quark star configuration when
compared to the case without diquark condensation. For finite
temperatures the masses are smaller than at $T = 0$.
For asymptotically high temperatures and densities
the EoS behaves like a relativistic ideal gas, where the relation
pressure versus energy density is temperature independent. In
contrast to the bag model where this behavior occurs immediately
after the deconfinement transition, our model EoS has a
temperature dependent $P(\varepsilon)$ relation also beyond this
point.

It has been shown for a hybrid star model which uses the quark
matter EoS presented in this work  that the possibility to obtain
a stable star configuration with 2SC quark matter core depends on
the formfactor of the quark interaction \cite{Blaschke:2003rg}.
The  Gaussian and  Lorentzian formfactor models do allow a quark
matter core, whereas the NJL formfactor model does not. Other NJL
model calculations \cite{Baldo:2002ju} confirm this conclusion.

A mass defect of about $0.1~M_{\odot}$ occurs in the cooling of a hot
configuration to a cold one already without diquark condensation.
It is mainly determined by the change of the configuration due to
changes in the EoS. Cooling the star with diquark condensation results
even in a lowering of the mass defect contrary to naive expectations.
For a more realistic calculation of the cooling evolution one has to start
from isentropic initial temperature profiles, which may change
the size of the mass defect but not our main conclusions.

\subsection*{Acknowledgements}

The authors are grateful to D.N. Aguilera and F. Sandin for discussions and
careful reading of the manuscript.
The work of H.G. was supported by DFG under grant No. 436 ARM 17/5/01.
D.B. and S.F. are grateful to the European Commission for support within the
Erasmus programme.

\newpage

\begin{table}[htb]
  \begin{tabular}{|c||c|c|c||c|c|}
 \hline
 Formfactor & $\Lambda$ & $G_1~\Lambda^2$ & $m$ & $T_c(\mu=0)$&$\mu_q^c(T=0)$\\
&[GeV]&&[MeV]& [MeV]& [MeV]\\
    \hline
    Gaussian & $1.0250$ & $3.761$ & $2.41$ & 175 &314\\
    Lorentzian  $\alpha = 2$ & $0.8937$ & $2.436$& $2.34$ & 188 & 330 \\
    NJL  & $0.9000$ & $1.944$& $5.10$ & 213 & 356\\
    \hline
  \end{tabular}
\vspace{1cm}
  \caption{Parameters for the different formfactors models discussed in the
text. Critical temperatures $T_c$ for the chiral phase transition at $\mu_q=0$
and critical chemical potentials  $\mu_q^c(T=0)$ for the onset of diquark
condensation at $T=0$.}
  \label{par}
\end{table}
\vspace{4cm}

\begin{table}[htb]
  \begin{tabular}{|c||c|c|c|c||c|c|c|c|}
    \hline
&\multicolumn{4}{c||}{Gaussian}&
\multicolumn{4}{c|}{Lorentzian}\\ \hline
 $T$& $M_{\rm max}$ & $R$ & $n_B(0)$ & $\varepsilon(0)$
    & $M_{\rm max}$ & $R$ & $n_B(0)$ & $\varepsilon(0)$\\
{\rm [MeV]}& $[M_{\odot}]$&[km]&$[n_0]$ & [GeV/fm$^3$]
           & $[M_{\odot}]$&[km]&$[n_0]$ & [GeV/fm$^3$]\\
    \hline
    $0^{*)}$ & $1.80$ & $10.10$& $ 6.97$ & $1.196$
             & $1.54$ & $ 8.72$& $10.34$ & $2.115$\\
    $0$   & $1.85$ & $10.00$& $ 7.54$ & $1.213$
          & $1.63$ & $ 8.63$& $ 9.24$ & $1.813$\\
    $40$  & $1.72$ & $ 9.67$& $ 7.70$ & $1.553$
          & $1.53$ & $ 8.44$& $ 9.81$ & $1.893$\\
    $60$  & $1.61$ & $ 9.46$& $ 7.95$ & $1.653$
          & $1.45$ & $ 8.55$& $ 9.97$ & $1.925$\\
    \hline
  \end{tabular}
\vspace{1cm}
  \caption{Characteristic quark star properties for Gaussian (left) and
Lorentzian (right) formfactor models: maximal possible masses $M_{\rm max}$,
radii $R$, central baryon densities $n_B(0)$ and central energy densities
$\varepsilon(0)$ for  different temperatures.\newline
$^*)$Diquark gap is set to $\Delta=0$.}
  \label{mass}
\end{table}

\end{document}